\documentclass[conference,a4paper]{IEEEtran}

 \usepackage{microtype}
\usepackage[utf8]{inputenc} 
\usepackage[T1]{fontenc}
\usepackage{url}              % provides \url{...}
\usepackage{ifthen}           % provides \ifthenelse
\usepackage{cite}             % improves presentation of citations
\usepackage[cmex10]{amsmath}  % Use the [cmex10] option to ensure compliance
\usepackage{adjustbox}
\usepackage{amssymb,amsfonts}
\usepackage{bbm}
\usepackage{dsfont}
\usepackage{mleftright}       % fix to wrong spacing of \left-,
\mleftright                   % \middle- \right-commands 
\usepackage{graphicx}         % provides \includegraphics{...} to
\usepackage{booktabs}         % fixes poor spacing in tables and
\usepackage{pgf}
\usepackage{caption}
\usepackage{lipsum} %for writing one-column equations in a two-column document with \begin{widetext}
\usepackage{cuted}
\usepackage{stfloats}
\usepackage{etoolbox}
\usepackage[switch]{lineno} 
\usepackage{setspace}

\newcommand{\pow}{\mathsf{P}}

\newcommand{\argmax}[1]{\ensuremath{\underset{#1}{\arg\max}}}
\newcommand{\sett}{\mathcal S}
\newcommand{\seto}{\Sigma}

\newcommand{\setd}{D^{\ka}\!}

\newcommand{\ka}{\ensuremath{\normalfont \textsf{K}_{\text{a}}}}

\newcommand{\dm}{\ensuremath{\textsf{D}_{\text{m}}}}

\newcommand{\textsfM}{{\normalfont \textsf{M}}}
\newcommand{\kaml}{\ensuremath{[\ka]\setminus \sett}}
\newcommand{\logM}{{\normalfont \log\textsf{M}}}

\newcommand{\T}{\ensuremath{t_0(\ba)}}
\newcommand{\ind}{\ensuremath{\mathbbm{1}}}
\newcommand{\ba}{\ensuremath{a^n}}

\newcommand{\ebno}{\ensuremath{\frac{\text{E}_{\text{b}}}{\text{N}_{\text{0}}}}}

\newtheorem{thm}{Theorem}
\newtheorem{lem}{Lemma}
\newtheorem{cor}{Corollary}

\newtheorem{defi}{Definition}

\newtheorem{rmk}{Remark}

\newenvironment{theorem}{\begin{thm}}{\vspace{0cm}\end{thm}}
\newenvironment{definition}{\begin{defi}}{\end{defi}}
\newenvironment{lemma}{\begin{lem}}{\end{lem}}

\newenvironment{remark}{\begin{rmk}}{\end{rmk}}
\begin{document}

\title{Worst-Case Per-User Error Bound for Asynchronous Unsourced Multiple Access } 

%%%%%%
\author{%
  \IEEEauthorblockN{Jyun-Sian Wu, Pin-Hsun Lin, Marcel A. Mross and Eduard A. Jorswieck}
  \IEEEauthorblockA{Institute for Communications Technology, Technische Universität Braunschweig, Germany\\
  \{jyun-sian.wu, p.lin, m.mross, e.jorswieck\}@tu-braunschweig.de
  % p.lin@tu-braunschweig.de
    }
  %   \and
  %   \IEEEauthorblockN{Pin-Hsun Lin}
  % \IEEEauthorblockA{Technical University of Braunschweig\\
  % p.lin@tu-braunschweig.de
  %   }
  %   \and
  %   \IEEEauthorblockN{Marcel Mross}
  % \IEEEauthorblockA{Technical University of Braunschweig\\
  % m.mross@tu-braunschweig.deC.G.F. }
  %   \and
  %   \IEEEauthorblockN{Eduard Jorswieck}
  % \IEEEauthorblockA{Technical University of Braunschweig\\
  % e.jorswieck@tu-braunschweig.de
  %   }
}

\maketitle

%%%%%
%% Abstract: 
%% If your paper is eligible for the student paper award, please add
%% the comment "THIS PAPER IS ELIGIBLE FOR THE STUDENT PAPER
%% AWARD." as a first line in the abstract. 
%% For the final version of the accepted paper, please do not forget
%% to remove this comment!
%%
\begin{abstract}
   This work considers an asynchronous $\ka$-active-user unsourced multiple access channel (AUMAC) with the worst-case asynchronicity. The transmitted messages must be decoded within $n$ channel uses, while some codewords are not completely received due to asynchronicities. We consider a constraint of the largest allowed delay of the transmission. The AUMAC lacks the permutation-invariant property of the synchronous UMAC since different permutations of the same codewords with a fixed asynchronicity are distinguishable. Hence, the analyses require calculating all $2^{\ka}-1$ combinations of erroneously decoded messages. Moreover, transmitters cannot adapt the corresponding codebooks according to asynchronicity due to a lack of information on asynchronicities. To overcome this challenge, a uniform bound of the per-user probability of error (PUPE) is derived by investigating the worst-case of the asynchronous patterns with the delay constraint. 
    Numerical results show the trade-off between the energy-per-bit and the number of active users for different delay constraints.  
    In addition, although the asynchronous transmission reduces interference, the required energy-per-bit increases as the receiver decodes with incompletely received codewords, compared to the synchronous case.
\end{abstract}

% !TeX root = ../ISIT2024.tex

\section{Introduction}
Internet-of-things (IoT), sensor networks, and ultra-reliable low latency massive machine-type communications have attracted attention for 6G communications and beyond. The main challenges of the codebook designs for these systems are: 1) short-blocklength codewords and 2) a large number of devices that an access point has to serve. Classical information theory uses the multiple-access channel (MAC) to analyze these systems. The classical MAC considers individual codebooks for all devices. However, the dramatically increasing number of devices prohibits using individual codebooks practically. 
In \cite{Polyanskiy_perspective_on_UMAC}, the author proposes a new system model, called \textit{unsourced multiple-access channel} (UMAC). For UMAC systems, all transmitters share an identical codebook, and the amount of data transmitted at each transmitter is the same.

There are several aspects to investigating UMAC. Authors in \cite{Guo_many_access_asymptotic_capacity} investigate the first-order capacity when the numbers of users are some functions of the blocklength, and users apply individual codebooks for identification and an identical codebook for transmitting information. The second-order asymptotic achievable rates of the grant-free random access system, where users access the channel without any prior request, are analyzed in \cite {Effros_RAC, Effros_RAC_MAC}. However, the achievable rates vanish if the number of transmitters increases asymptotically. Therefore, authors in \cite{Polyanskiy_perspective_on_UMAC} investigate the energy efficiency of synchronous UMAC with per-user error probability (PUPE) constraint. 
Authors in \cite{T_fold_1} propose the T-fold ALOHA and a low-complexity coding scheme for the grant-free Gaussian random access channel, where the coding scheme is based on the compute-and-forward \cite{Compute-and-forward} scheme and coding for a binary adder channel. Authors in \cite{T_fold_1} also analyze the energy efficiency of the T-fold ALOHA and the low-complexity coding scheme.
% Synchronizing all devices is a difficult task. Therefore, asynchronous systems need to be considered as well. 

Asynchronous systems are worth investigating due to the difficulty of synchronizing a large number of devices.
For asynchronous classical MAC, the capacity is the same as the synchronous MAC \cite{AMUC1981}, assuming the ratio of delay to blocklength asymptotically vanishes. For asynchronous UMAC (AUMAC), authors in \cite{Polyanskiy_low_complexity_AUMAC, Polyanskiy_Short_packet} utilize the T-fold ALOHA \cite{T_fold_1} and the orthogonal frequency-division multiplexing (OFDM), transforming the time-asynchronous problem to a frequency-shift problem. The maximum delay in \cite{Polyanskiy_Short_packet} must be smaller than the length of the cyclic prefix. Authors in \cite{Guo_CS} apply a sparse orthogonal frequency-division multiple access (OFDMA) scheme and compressed sensing-based algorithms to reliably identify arbitrarily asynchronous devices and decode messages. 

We consider the AUMAC system with a bounded delay, i.e., maximum delay $\dm\in \mathbb Z^+\cup 0$, and $\frac{\dm}{n}$ is a constant w.r.t. the blocklength $n$. Transmitters transmit a fixed payload size with an identical finite-length $n$ codebook. The delays of active users are smaller than $\dm$. 
In our considered model, the messages have to be decoded within $n$ channel uses. Receivers decoding without completely receiving codewords are investigated in broadcast channels \cite{PH_ED1, MM_ED2}. We analyze the PUPE of AUMAC with decoding from incompletely received codewords while assuming the blocklength is finite. 
To provide a more precise analysis than the typically used Berry-Esseen theorem (BET) in finite blocklength analyses \cite{saddlepoint_of_RCUs}, we apply the saddlepoint approximation \cite{Saddlepoint_approximation_keystone}. 
% This work applies saddlepoint approximation \cite{Saddlepoint_approximation_keystone} to analyze the PUPE. 
In the synchronous UMAC, for any $1\leq k\leq \ka$, all combinations that $k$ out of $\ka$ messages are decoded erroneously have identical tail probabilities due to the permutation-invariant property. However, the permutation-invariant property is invalid due to the asynchronicity. In particular, each $k$ out of $\ka$ combination of the erroneously decoded messages has a different tail probability, while $k\in[\ka]$.
Therefore, the analysis requires the sum of $2^{\ka}-1$ different tail probabilities. To overcome this computational challenge, we derive a uniform upper bound of PUPE for our considered AUMAC. This bound allows us to: 1) analyze the PUPE without calculating every combination of the erroneously decoded messages and 2) evaluate the required energy to satisfy the PUPE constraint and transmit the payload.
Analyses show that even though the AUMAC has less interference than synchronous UMAC, the reduction of PUPE due to the increasing number of received symbols is more significant than the increment of the PUPE due to the interference.
Numerical results compare achievable energy efficiencies for the proposed AUMAC to synchronous UMAC, which can be considered a special case of AUMAC with $\dm=0$. Numerical results show that compared to synchronous UMAC, transmitters in AUMAC require more energy to reliably transmit messages with a constant $\frac{\dm}{n}$. 

\textit{Notation:} We will denote $f^{(i)}(t)$ as the $i$-th derivative of $f(x)$ at the point $x=t$ and $f_{1,y}^{(i)}(x,t)$ as the $i$-th partial derivative of $f_1(x,y)$ w.r.t. $y$ at the point $y=t$. We use the indicator function $\ind(\cdot)$, the natural logarithm $\log(\cdot)$, and the Landau symbol $O(\cdot)$. The binomial coefficient of $n$ out of $k$ is represented by $\binom{n}{k}$. The number of permutations of $k$ is denoted as $k!$. We define $j=\sqrt{-1}$. We denote $[k]=\{1,2,...,k\}$ and $\mathcal F\setminus\mathcal T=\{x:x\in \mathcal F, x \not\in\mathcal T\}$, where $\mathcal F$ and $\mathcal T$ are two sets. We also denote $\mathbb Z^+_0=\mathbb Z^+\cup 0$. For any set $\mathcal F=\{F_1,F_2,...,F_{|\mathcal F|}\}$, we denote $\{X_m\}_{m\in \mathcal F}=\{X_{F_1},X_{F_2},...,X_{F_{|\mathcal F|}}\}$.
% !TeX root = ../ISIT2024.tex
% \vspace{-0.2cm}
\section{System model and preliminaries} \label{sec: system model} \vspace{-0.5mm}
We consider an AUMAC, which has additive white Gaussian noise (AWGN), one receiver, and multiple transmitters, where the number of active transmitters is denoted by a positive integer $\ka$. All transmitters utilize the same codebook with the same maximal power constraint, $\pow'$, to transmit the same (and fixed) size of payloads, i.e., $\logM$ nats, to the receiver. The codewords are independent and identically distributed (i.i.d.) generated from a Gaussian distribution with mean zero and variance $\pow$, where $\pow<\pow'$ due to the power backoff. The power backoff reduces the probability that the maximal power constraint violations occur.

\begin{definition}\label{def: delay set}
    We define the asynchronicity in terms of the vector of time shifts (delay) as 
    \vspace{-1.2mm}
    \begin{align*}
    \setd:=[d_1,\ d_2,\ ..., \ d_{\ka }] \in \{\mathbb Z^+_0\}^{\ka},
    \vspace{-2.5mm}
    \end{align*}
    % \vspace{-1mm}
    where $0=d_1,\; d_i\leq \dm$ and $ d_i\leq d_\ell,\; \forall \ell> i$ for all $i\in[\ka]$. The $i$-th entry, $d_i$, represents the delay of the $i$-th received codeword relative to the first received codeword, and $\dm$ denotes the delay constraint. We define $\alpha:= \frac{\dm}{n}\in[0,1)$, which is constant w.r.t. the blocklength $n$, and $\bar \alpha=1-\alpha$. 
\end{definition}

We assume that the receiver has perfect knowledge of the asynchronicity \cite{Csiszar_AMAC} and jointly detects the transmitted messages.
Asynchronous communication systems may result from asynchronous clocks between transmitters and receivers, different idle times among transmitters, or channel delays. 

\begin{remark}
    We consider that every transmitter transmits with the same codebook, and the receiver is not interested in identifying the senders of the received codewords. Therefore, $d_i$ indicates the delay of the $i$-th received codeword but does not indicate the identification of the transmitter. 
\end{remark}

\begin{figure}[htbp]\vspace{-4mm}
  \centering
  \includegraphics[width=0.28\textwidth]{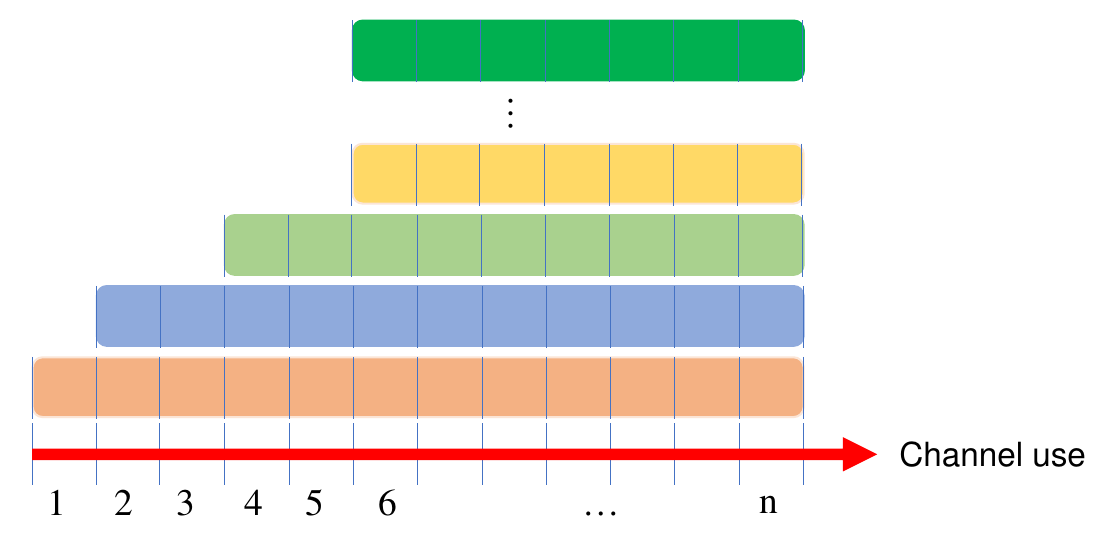}
  \caption{A $\ka$-active-user AUMAC with $\setd=[0,1,3,5,...,5]$. 
  % {\color{red}I'm afraid this example is too simple and then can be misleading: we consider many transmitters but here there are only 4. Maybe we can consider Ka=500, but in your example, you just choose the first few time samples for S and the figure can be with "..." to show there are really many transmitters}
  }
  \label{fig: AUMAC example}
\end{figure}

In the asynchronous model, the number of transmitted codewords symbols of each channel use can be different. For a given delay $\setd$ and the set of erroneously decoded messages $\sett \subseteq [\ka]$, we define a vector 
\vspace{-1mm}
\begin{flalign}
\ba(\sett,\,\setd)\!:=\![a_1(\sett,\,\setd),a_2(\sett,\,\setd),\!...,a_n(\sett,\,\setd)], \label{eq:def a}
\end{flalign}
where $a_i(\sett,\setd)\leq a_\ell(\sett,\setd)$, $\forall \ell >i$, $i\in[n]$ and $a_i(\sett,\setd)\in \mathbb Z^+_0$, $\forall i\in[n]$. For a given $\setd$ and a given $i\in[n]$, the $i$-th entry of $\ba(\sett,\setd)$, i.e., $a_i(\sett,\setd)$, indicates the number of simultaneously received symbols, which belong to $\sett$, at the $i$-th channel use. 
To simplify notations, we use $\ba:=[a_1,a_2,...,a_n]$ to represent $a^n(\sett,\setd)$. For example, considering a $\ka$-active-user AUMAC with $\setd=[0,1,3,5,...,5]$ in Fig. \ref{fig: AUMAC example}, for the set $\sett\!=\!\{1,2\}$, $\ba\!=\![a_1\!=\!1,a_{[n]\setminus[1]}\!=\!2]$; for the set $\sett\!=\!\{2,3,4\}$, $\ba\!=\![a_1\!=\!0,a_2\!=\!1,a_3\!=\!1,a_4\!=\!2,a_5\!=\!2,a_{[n]\setminus[5]}\!=\!3]$. Note that for a given $\sett$ and $\setd$, $a_{[n]\setminus [\alpha n]}=|\sett|$.

\vspace{-1.5mm}

For any $\ell\in[n]$, we define a shift function $\tau_{d_i}(X^n_{i},\ell):=X_{i,\ell-d_i}$, where $X_{i,\ell-d_i}$ is the $(\ell-d_i)$-th element of $X^n_i$, and if $\ell-d_i\not \in[n]$, $X_{i, \ell-d_i}=0,\ \forall i\in[\ka]$. 
The received symbol at the receiver at time $\ell\in [n]$ is
\vspace{-1mm}
\begin{align}
	 {Y}_\ell = \sum_{i=1}^{\ka }\tau_{d_i}\left(X^n_i,\ell\right)+  Z_\ell, \label{eq:system_model2}\vspace{-2mm}
\end{align}
where the channel input $X^n_i \!\in \!\chi ^n\!\subset\! \mathbb R^n$, where 
% \begin{align}
    $\chi^n\!:=\!\{x^n :\! x^n\! \in \mathbb R^n,\; \|x^n\|^2\!\leq\! n\pow'\}$
% \end{align}
is the channel input satisfying the maximal power constraint and $Z_\ell\sim \mathcal N(0, 1)$ is an i.i.d. AWGN, $\forall \ell\in[n]$.
% We define the joint distribution $P_{X_{[\ka]}}(\taud (x^n_{[\ka]})_m)=\Pi\limits_{\ell:\ \ell\in[\ka],\ m-d_\ell\in[n]} P_X()$ 
% Fig. \ref{fig: the illustration of the received signal of AUMAC} illustrates the received signal in the receiver. 

\begin{definition}\label{definition1}
    An $(n,\textsfM, \epsilon, \ka, \alpha, \setd)-$code, $\mathcal C_1$, for an AUMAC described by $P_{Y|X_{[\ka]}}$, consists of
    \begin{itemize}
    \item one message set $\mathcal M=\{1,2,...,\textsfM\}$,
    \item one encoder $f: \mathcal M \to \chi^n$,
    \item one decoder $g: \mathbb R^n \to\binom{[\mathcal M]}{\ka},$ where $\binom{[\mathcal M]}{\ka} $ is a set containing $\ka$ distinct elements from the set $\mathcal M$,
    \end{itemize}
    % such that all codewords fulfill the maximal power constraint $\|x^n_i\|^2\leq n \pow'$, $\forall i\in[\ka]$, 
    and the delay $\setd$ fulfills the delay constraint $\alpha n$ in Def. \ref{def: delay set}, the PUPE satisfies
    \vspace{-1.5mm}
    \begin{align}
        % \max\limits_{\setd: d_i\leq \alpha n,\ \forall i\in[\ka]}
        % {\mbox{P}_{\text{PUPE}}}:=\max\limits_{\setd:\ d_{\ka}\leq \alpha n}\sum_{i=1}^{\ka}\frac{1}{\ka}\mbox {Pr}(E_i|\setd)\leq \epsilon,
        {\mbox{P}_{\text{PUPE}|\setd}}:=\frac{1}{\ka}\sum_{i=1}^{\ka}\mbox {Pr}(\tilde {\mathcal E}_i|\setd)\leq \epsilon,\vspace{-2.5mm}
    \end{align}
    where $ \tilde{\mathcal E}_i  :=  \{\;  \cup_{\ell \neq i}\{M_i  = M_\ell \} \cup \{M_i \not\in  g({Y}^n) \} \cup \{\|f(M_i)\|^2 >  n\pow'\} \}$,  $i   \in   [\ka]$, and $M_i  \sim  \mbox{Unif} (\mathcal M)$ is the $i$-th transmitted message.
\end{definition}

% !TeX root = ../ISIT2024.tex

\section{Main Results}
% \subsection{Achievable Result}

There are several achievable schemes in the synchronous multiple-access models. For synchronous UMAC models, shell codes achieve better second-order asymptotic rates than the i.i.d. Gaussian codes \cite{Effros_RAC_MAC, Effros_RAC}. However, in our considered asynchronous model, the receiver decodes the messages solely based on the first $n$ received symbols, i.e., some codewords are incompletely received when the receiver starts to decode. The decoding based on incompletely received codewords does not satisfy the definition of shell codes. This fact prevents us from analyzing the models by the uniform distribution on a power shell. Therefore, we consider an achievable scheme that all transmitters share the same codebook i.i.d. generated by a  Gaussian distribution $P_X$. 
% We define the information density $i(X^n;Y^n)=\log\frac{d P_{Y^n|X^n}}{d P_{Y^n}}$. 
The receiver performs maximum information density decoding with knowledge of $\setd$, which is defined by \vspace{-1mm}
\begin{align}g(Y^n)= \argmax{X^n_{[\ka]}\in \mathcal C_1}\ \sum_{\ell=1}^{n}i\left(\big\{\tau_{d_m}(X^n_{m},\ell)\big\}_{m\in[\ka]}; Y_\ell\right),\label{eq: def. of decoder}\vspace{-1mm}\end{align}
where
\begin{align}
i\Big(\big\{\tau_{d_m}(X^n_{m},\ell)&\big\}_{m\in[\ka]}; Y_\ell\Big):=\nonumber\\
&\log\left(\frac{dP_{Y|X_{[\ka]}}\left(Y_\ell|\big\{\tau_{d_m}(X^n_{m},\ell)\big\}_{m\in[\ka]}\right)}{dP_{Y}(Y_\ell)}\right).\nonumber
\end{align}
% and
% \begin{align*}
%     &P_{X_{[\ka]}}\left(\big\{\taud(X^n_{m},\ell)\big\}_{m\in[\ka]}\right)\\
%     &\hspace{3cm}=\prod\limits_{m\in[\ka]: \ \ell-d_m\in[n]} P_{X}(\taud(X^n_m,\ell)).
% \end{align*}

The corresponding finite-blocklength (FBL) analysis results are summarized in the following.
\begin{theorem}
\label{thm1}
    Fix $0<\pow<\pow'$. There exists an $(n,\textsfM,\epsilon,\ka,\alpha,\setd)-$code for an AUMAC such that the PUPE can be upper bounded by the following:
    \begin{flalign}
        &\sum\limits_{\sett\subseteq[\ka]}\!\!\frac{|\sett| g_1(\!\ba\!,\T) }{\ka\sqrt{2\pi}} \!\Big[g_2(\ba\!,\! \T)\!\!+\!\xi(\ba,\!\T)\!\Big]\!\!+\!p_0\!\!\leq\! \epsilon, 
        \label{eqthm1: main result}\vspace{-1mm}
    \end{flalign}
    % where $\T\in \mathbb R$ satisfies $E^{(1)}_t(\ba, \T)=|\sett|\logM$,
    if there exists a $\T\in(0,1)$ such that $E^{(1)}_t(\ba, \T)=|\sett|\logM$, where 
    % $\mathcal R:=\{t: t\in(0,1)\}$, 
    \begin{flalign}
        g_1(\ba,t):=&\exp(t |\sett|\logM-E(\ba,t)),\label{eqthm1: defg1}\\
        g_2(\ba,t):=&\left(t(1-t)\sqrt{-E_t^{(2)}(\ba,t)}\right)^{-1},\label{eqthm1: defg2}\\
        E(\ba,t ):=&\frac{1}{2}\sum\limits_{i=1}^{n}\left(\!t\log(1\!+\!a_i \pow)\!+\!\log\left(1\!-\!\frac{a_i \pow t^2}{1\!+\!a_i \pow}\right)\right), \label{eq: def: E(a,t)}\\
        \xi(\ba,t):=&\frac{1}{2\pi j}\int^{t+j\infty}_{t-j \infty}\exp\left(-\frac{E^{(2)}_{t}(\ba,t)}{2}(\rho-t)^2\right) \nonumber\\
        &\hspace{1.5cm}\cdot\frac{1}{\rho(1-\rho)}\sum_{m=1}^{\infty}\frac{\bar \xi(\ba,t,\rho)^m}{m!} d\rho ,\!\!\label{eq:def of high order term in Taylor seriers}\\
    % \end{align}
    % \begin{align}
        \bar \xi(\ba,t,\rho):=&-\sum_{i=3}^{\infty} E^{(i)}_t(\ba, t)\frac{(\rho-t)^{i}}{i!},\vspace{-1mm}
        \end{flalign}
    and $p_0:=\frac{\ka(\ka-1)}{2\textsfM}+\sum\limits_{i=1}^{\ka }\mbox{Pr}(\|X^n_i\|^2> n\pow').$
    % \label{eq:p0}
    % \color{red} We adjust $\frac{B_{\ba}}{\sqrt{\bar \alpha n}}$, which is from large deviation theorem bound [], to our asynchronous model. \color{black}
    % \begin{align}\label{eq:def of Ba}
    %     \frac{B_{\ba}}{\sqrt{\bar \alpha n}}=...
    % \end{align}
\end{theorem}

 In our PUPE analysis, two main tools are the Taylor expansion and the inverse Laplace transform. The former tool is used to expand the exponent of $g_1(\ba,t)$ at $t=\T$ and the latter one is used to derive the probability density function (PDF) from the cumulant-generating function (CGF), where CGF is the logarithm of the moment-generating function (MGF). The sum of higher order terms of the Taylor expansion at $t=\T$ is represented by $\bar \xi(\ba,\T,t)$. The proof is relegated to Appendix \ref{app:sec:thm1 proof}.
    % \color{red} the contents that are relevant to LDT bound will be replaced by Chernoff bound\color{black}

Theorem \ref{thm1} can evaluate the PUPE for any delay $\setd$ satisfying the delay constraint $\alpha n$. However, evaluating (\ref{eqthm1: main result}) requires calculating all $\sett\subseteq [\ka]$, which is infeasible if $\ka$ is sufficiently large.
    % More specifically, many codebooks are designed for different $\setd$'s, 
% \begin{remark}
%     For the case $\mbox{Pr}(M_1 \not \in g(Y^n))=\mbox{Pr}\left(\bigcup_{\sett\subseteq[\ka], 1\in \sett} M_{\sett}\not \in g(Y^n)\right)$. Therefore, evaluating the error probabilities of all possible error events is necessary when considering PUPE instead of joint error probability.
% \end{remark}
Additionally, even though we can design different codebooks for different $\setd$ satisfying the PUPE constraints, the transmitters cannot select the codebook corresponding to a particular $\setd$ since they have no information on delays. 
Therefore, a codebook that satisfies the PUPE constraint regardless of $\setd$ is required.
In the following, we derive a uniform upper bound of the PUPE for all $\setd$'s satisfying delay constraint $\alpha n$.
     
\begin{definition}\label{definition 2}
    An $(n,\textsfM, \epsilon, \ka, \alpha)-$code, $\mathcal C_2$, for an AUMAC described by $P_{Y|X_{[\ka]}}$ consists of one message set $\mathcal M$, one encoder $f$, and one decoder $g$ defined by 
    \begin{align}
        g(Y^n)= \argmax{X^n_{[\ka]}\in \mathcal C_2}\ \sum_{\ell=1}^{n}i\left(\big\{\tau_{d_m}(X^n_{m},\ell)\big\}_{m\in[\ka]}; Y_\ell\right),\vspace{-1mm}
    \end{align}
    such that for the power constraint $\pow'$ and any $\setd$ satisfying the maximum delay constraint, the PUPE satisfies
    \begin{align}
        {\mbox{P}_{\text{PUPE}}}:=\max\limits_{\setd:\ d_{\ka}\leq \alpha n}\sum_{i=1}^{\ka}\frac{1}{\ka}\mbox {Pr}(\tilde {\mathcal E}_i|\setd)\leq \epsilon,\vspace{-1mm}
    \end{align}
    where $\tilde {\mathcal E}_i$ is defined in Definition \ref{definition1}.
\end{definition}
% \color{red} supplement decoder definition \color{black}

Based on the PUPE defined in Def. \ref{definition 2}, we find the $a^{*n}_\iota$ that leads to the uniform upper bound of the PUPE, where $\iota\in\{0,1\}$. 

\begin{theorem}
\label{thm2}
Fix $0<\pow< \pow'$. There exists an $(n,\textsfM,\epsilon,\ka,\alpha)-$code for AUMAC, such that the PUPE can be upper bounded by the following:
    \begin{align}
    \label{eq:thm2}
        &\frac{1}{\ka\sqrt{2\pi}}\sum_{|\sett|=1}^{\ka} \left(\binom{\ka-1}{|\sett|}\frac{|\sett| g_1(a_0^{n*},t_0(a_0^{n*}))}{T_0^*\sqrt{-E^{(2)}_t(a_0^{n*}, \underline t_0)}}\right.\nonumber\\
        &\!\!+\!\!\hspace{0cm}\left.\binom{\ka-1}{|\sett|-1}\frac{|\sett| g_1(a_1^{n*},t_0(a_1^{n*}))}{T_1^*\sqrt{-E^{(2)}_t(a_1^{n*}, \underline t_1)}}\! \right)\!\!+\!p_0\!+\!O\left(\frac{\exp(-n)}{\sqrt{n}}\right)\!\leq\epsilon ,\vspace{-1mm}
    \end{align}
    if $t_0(a^{n*}_\iota)\in\mathcal A\cap \mathcal B ,\ \bar t_\iota \in\mathcal A\cap \bar{\mathcal B}$, $\ \underline t_\iota \in\mathcal A\cap \underline{\mathcal B}$, and $\underline t_\iota \leq \T\leq \bar t_\iota$, where $a_\iota^{n*}=[\iota^{\alpha n}, |\sett|^{n-\alpha n}]$, $T^*_\iota:=\min\{\underline t_\iota-\underline t_\iota^2,\ \bar t_\iota-\bar t_\iota^2\}$, $\iota:=\ind(1\in\sett)$, $\mathcal {A}\!:=\!\left\{t\!:\!t\in(0,1) \right\}$,
    \begin{flalign}
        &\mathcal B\!:=\!\left\{t\!:\!E^{(1)}_t(a^{n*}_\iota, t)=|\sett|\logM\right\},\label{def: t}\\
        &\bar{\mathcal B}\!:=\!\left\{\!t\!:\!\!\sum_{i=1}^{n}\!\frac{a^*_{\iota,i}\pow t}{\!1\!\!+\!a^*_{\iota,i}\pow(1\!-\! t^2)\!}\!\!=\!\!\frac{n}{2} \!\log(1\!\!+\!|\sett|\pow)\!\!-\!|\sett|\logM\!\right\},\label{def:bar t}\!\!\!\!\!\\
        &\underline{\mathcal B}\!:=\!\left\{t\!:\!\frac{|\sett| n\pow t}{\!1\!\!+\!|\sett|\pow(1\!-\!t^2)\!}\!=\!\!\sum_{i=1}^{n}\!\frac{\log(1\!\!+\!a^*_{\iota,i}\pow)}{2}\!-\!|\sett|\logM\!\right\},\label{def:underline t}\!\!\!\!\\
        &\text{and } a^{*}_{\iota,i} \text{ is the } i \text{-th element of } a^{n*}_\iota.\nonumber
    \end{flalign}
    \vspace{-4mm}
\end{theorem}
The proof is relegated to Appendix \ref{app:sec:thm2 proof}.
% The upper bound and the lower bound of $\T$ exist for any $\ba$, denoted as 
% $\bar t_\iota$ and $\underline t_\iota$, respectively. If the corresponding $\sett$ of $\ba$ contains the first arrived codeword, i.e., $1 \in \sett$, then $\iota=1$ since the assumption that the first arrived codeword has delay $0$ in our considered model, i.e., the minimum of $a^*_{1,[\alpha a]}=1$. Otherwise, $\iota=0$ and the minimum of $a^*_{0,[\alpha n]}=0$. In other word, for a fixed $\setd$, every $\sett$ influences $\underline t_\iota$ and $\bar t_\iota$ by the cardinality $|\sett|$ and whether $1 \in \sett$.
% Therefore, the uniform PUPE upper bound introduced in Theorem \ref{thm2} are functions of $a^*_0, T^*_0, \bar t_0$ and $\underline t_0$, and $a^*_1, T^*_1, \bar t_1$ and $\underline t_1$ for $1\not \in\sett$ and for $1\in\sett$, respectively. 
% % , \color{red}i.e., $\underline t_\iota$ and $\bar t_\iota$ are constant for the same . \color{black}

The benefit of having a uniform PUPE upper bound is that it allows us to analyze the performance of an $(n,\textsfM,\epsilon,\ka,\alpha)-$code without calculating all tail probabilities of the corresponding possible $\sett$ but scaling the uniform PUPE upper bound by a binomial coefficient.

The term $g_1(\ba,\T)\xi(\ba,\!\T)$ in Theorem \ref{thm1} is expressed by $O\left(\frac{\exp(-n)}{\sqrt{n}}\right)$ in Theorem \ref{thm2} since $\bar \xi(\ba,t,\rho)$ behaves as $O(n^{-\frac{1}{2}})$ \cite{Saddlepoint_approximation_keystone}.
We refer to both the approximations obtained by ignoring $g_1(\ba,\T)\xi(\ba,\T)$ and $O\left(\frac{\exp(-n)}{\sqrt{n}}\right)$ in (\ref{eqthm1: main result}) and (\ref{eq:thm2}), respectively, as saddlepoint approximations. 
The saddlepoint approximation has an exponentially decreasing approximation error w.r.t. $n$, allowing us to obtain sufficiently precise approximations of the FBL PUPE compared to the BET.

\begin{remark} 
\label{remark: reducing a increases error prob}
The upper bound in Theorem \ref{thm1} decreases as $a_i$ increases for any $\setd$ and $\sett$ because $\frac{\partial}{\partial a_i} g_1(\ba,t)g_2(\ba,t)\leq 0$ for all $t\in(0,1)$ and $i\in[\alpha n]$. 
In fact, having more overlap in the transmission leads to more interference. 
% However, the analytical results show that the reduction of PUPE due to the increasing number of received symbols is more significant than the increment of the PUPE due to the interference.
A larger number of overlapping symbols has one positive and one negative effect on the receiver: it leads to more received energy but, meanwhile, more interference. By our analysis, we found that the positive effect is dominant.
\end{remark}

% !TeX root = ../ISIT2024.tex

\section{Numerical Results}
\begin{figure}[htbp]
  \centering
  \includegraphics[width=0.35\textwidth]{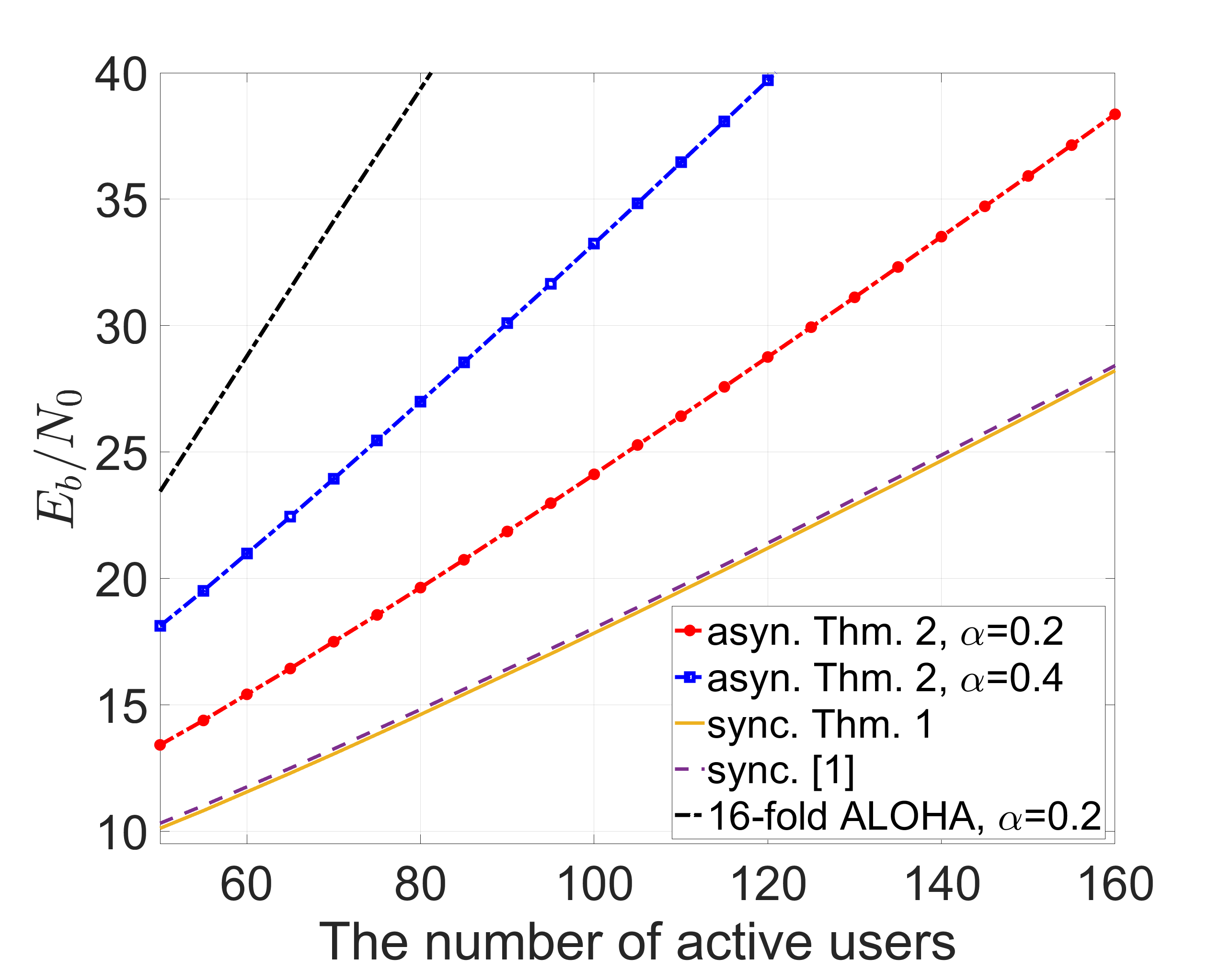}
  \caption{$\ebno$ of AUMAC compared to synchronous UMAC for different numbers of active users.}
  \label{fig: AUMAC result}
  \vspace{-2.5mm}
\end{figure}

Based on the results of Theorem \ref{thm1} and Theorem \ref{thm2} without considering $g_1(\ba,\T)\xi(\ba,\T)$ and $O\left(\frac{\exp(-n)}{\sqrt{n}}\right)$, we numerically evaluate the energy-per-bit versus the number of active users. 
We define the energy-per-bit as $\ebno:=\frac{n\pow'}{\logM}$. The PUPE upper bounds from Theorem \ref{thm2} are compared to two UMAC schemes under different scenarios with the following parameters: $\logM=100$, $n=4000$, $\epsilon=10^{-3}$ and $\ka\in[50, 160]$. 
In Fig. \ref{fig: AUMAC result}, the purple dash curve of \cite{Polyanskiy_perspective_on_UMAC} is evaluated by numerically optimizing $\pow<\pow'$. We also evaluate the $\ebno$ for a $16$-fold ALOHA \cite{T_fold_1} with Theorem \ref{thm1}. The $16$-fold ALOHA splits the transmission into $V$ subblocks such that the collision probability is less than $0.9 \epsilon$ \cite{T_fold_1}. Each subblock has blocklength $\tilde n=n/V$ and the delay constraint for each subblock is $\alpha \tilde n$. We assume that the messages have to be decoded within $\tilde n$ channel uses. The black dot-dash curve shows that the required $\ebno$ for $16$-fold ALOHA from Theorem \ref{thm1}.
The synchronous UMAC can be considered to be a special case of AUMAC with $\alpha =0$.
% The yellow dot-dash curve indicates that the required $\ebno$ of the synchronous UMAC can be considered to be a special case of
Additionally, analyzing synchronous UMAC does not require calculating the number of permutations of erroneously decoded messages, i.e., $|\sett|!$, so we modify (\ref{eqthm1: defg1}) in Theorem $\ref{thm1}$ as follows:
\begin{align}
g_1(\ba,t):=&\exp\left(t \log \left( \binom{\textsfM-\ka}{|\sett|}\right)-E(\ba,t)\right). \label{eq: sync thm1}
\end{align} 
% The yellow dot-dash curve indicates that the required $\ebno$ of the synchronous UMAC evaluated from Theorem $\ref{thm1}$ with modification \eqref{eq: sync thm1}.
The yellow solid curve shows the required $\ebno$ for the synchronous UMAC. It is computed by numerically optimizing $\pow<\pow'$ in Theorem \ref{thm1} with $\setd=\{d_1=d_2=...=d_{\ka}=0\}$ and adapting \eqref{eq: sync thm1}.

We numerically optimize $\pow$ in Theorem $\ref{thm2}$ with $\alpha=0.2$ and $\alpha=0.4$ and compare the $\ebno$ of AUMAC and that of synchronous UMAC. 
Numerical results show that the AUMAC that has larger $\alpha$ causes the transmitters to consume more energy to transmit in the worst case of delay. Observing the curves of $\alpha =0.2$, $\alpha=0.4$ and $\alpha=0$ (synchronous), we can conclude that for the AUMAC with larger $\alpha$, which means fewer interference for the first $\alpha n$ channel uses, the PUPE increases. It is because the receiver decodes the messages based on fewer transmitted codewords symbols, which is equivalently based on less received energy. This effect is illuminated in Remark \ref{remark: reducing a increases error prob}. Thus, codebooks of our considered model require more energy to achieve the same PUPE constraint. 
% {\color{red} Any comments on the gap of the UB? Is the gap analyzable or parameterizable?}
\vspace{-0.3cm}

% !TeX root = ../ISIT2024.tex

\section{Conclusions}

In this work, we analyze the FBL performance of the asynchronous UMAC system with bounded and non-vanishing delay constraints $\alpha n$. The derivations based on the saddlepoint approximation provide FBL performance bounds. We also investigate a uniform upper bound of the PUPE, which highly simplifies the analysis to multiply the uniform upper bound with the corresponding binomial coefficient instead of calculating tail probabilities of all error events. The numerical results show the trade-off between $\ebno$ and delay constraint $\alpha n$. Although asynchronous transmissions have less interference, reducing the error probability of the first few codewords, it increases PUPE as the receiver decodes shorter codewords, which is analytically shown in Theorem $\ref{thm2}$ and is numerically shown in our numerical results. Compared to the synchronous case, the achievable energy-per-bit $\ebno$ for the asynchronous case shows that the required $\ebno$ increases as the receiver decodes shorter codewords, even though interference reduces.
\newpage
\bibliographystyle{IEEEtran}
\enlargethispage{-12cm}
\IEEEtriggeratref{10}
% \IEEEtriggeratref{4}
\bibliography{references}

%%
%% where we here have assumed the existence of the files
%% definitions. Bib and bibliofile. Bib.
%% BibTeX documentation can be obtained at:
%% http://www.ctan.org/tex-archive/biblio/bibtex/contrib/doc/
%%%%%%
%% Or you use manual references (pay attention to consistency and the
%% formatting style!):

%%%%%% 
%% Appendix:
%% If needed, a single appendix is created by
%%
% \newpage
\clearpage
\appendices
% !TeX root = ../ISIT2024.tex

\section{Proof of Theorem 1}\label{app:sec:thm1 proof}

Theorem 1 is derived by the maximal information density decoder with the random coding union (RCU) bound \cite{Polyanskiy.2010} to express the per-user probability of error (PUPE) as a sum of tail probabilities. 
% The Taylor expansion and the inverse Laplace transform evaluate the tail probabilities precisely. 
In the following, we first show the expressions of PUPE regarding tail probabilities. Then, we apply the Taylor expansion and the inverse Laplace transform to derive the tail probabilities.

We define 
% $E_i := \{M_i =M_\ell, \forall \ell\neq i\}$, $i\in [\ka]$, which represents that the transmitter, indexed by $i$, transmits a non-unique message,
$\tilde {\mathcal E}_\ell:=\{\;\cup_{i\neq \ell}\{M_\ell =M_i\}\cup\{M_\ell\not\in g({Y}^n )\}\cup \{\|f(M_\ell)\|^2> n\pow'\}\}$, which represents the $\ell$-th user's error event of the PUPE, ${\mathcal E}_\ell: =\{\{M_\ell \neq M_i,\;\forall i \neq \ell\}\cap \{\|f(M_i)\|^2\leq n\pow',\;\forall i \in [\ka]\}\}$, $\ell\in [\ka]$, which represents the event that the other transmitted messages are distinct to the $i$-th transmitted message and transmitted codewords fulfill the power constraint, and $p_0:=\frac{\ka(\ka-1)}{2\textsfM}+\sum\limits_{i=1}^{\ka }\mbox{Pr}(\|X^n_i\|^2>n\pow')$ is the upper bound of the probability that collisions or power constraint violations occur. 

The PUPE of an $(n, \textsfM, \epsilon, \ka,\alpha, \setd)$-code can be upper bounded by the union bound as follows:
\begin{align}
        {\mbox{P}_{\text{PUPE}|\setd}}:&=\sum_{\ell=1}^{\ka}\frac{1}{\ka}\mbox {Pr}\Big(\tilde {\mathcal E}_\ell|\setd\Big)\\
        &\leq p_0+\sum_{\ell=1}^{\ka}\frac{1}{\ka}\mbox {Pr}\Big(M_\ell \not \in g(Y^n)|\setd, {\mathcal E}_\ell\Big).\label{apen1: proof thm1; split to no-repetition event}
\end{align}

To simplify the notation, we omit the condition $\setd$ in the following derivation. 
For any subset $\sett\subseteq [\ka]$, we define 
\begin{align*}
&\tilde \gamma(\bar X^n_\sett, X^n_{[\ka]\setminus \sett})\\
&\hspace{0.8cm}\!\!:=\!\!\sum_{\ell=1}^{n}i(\{\tau_{d_m}(\bar X^n_{m},\ell)\}_{m \in \sett},\!\{\tau_{d_m}(X^n_{m},\ell)\}_{m\in \kaml};Y_\ell)\\ 
&\text{and} \\
&\gamma(\bar X^n_\sett)\!\!:=\!\!\sum_{\ell=1}^{n}\!i(\{\tau_{d_m}(\bar X^n_{m},\ell)\}_{m \in \sett};\!Y_\ell|\{\tau_{d_m}( X^n_{m},\ell)\}_{m \in \kaml}).
\end{align*}
We define a set 
\begin{align}
    \seto(\ell):=\{\sett: \sett\subseteq [\ka], \ell\in \sett\}, \label{app: eq: def of seto}
\end{align}
which contains all possible subsets $\sett$ of the error event $\{M_\ell \not \in g(Y^n)\}$.
% , and $$\Omega :=\bigcup_{\ell \in [\ka]} \Omega(\ell)=\{\sett:\sett\subseteq [\ka]\}.$$  
Substitute the definition of the maximal information density decoder into $\mbox {Pr}(M_\ell \not \in g(Y^n)|{\mathcal E}_\ell)$, we have
% (\ref{eq: def. of decoder}) into $\mbox {Pr}(M_\ell \not \in g(Y^n)|{\mathcal E}_\ell)$. We have
\begin{flalign}
    &\mbox{Pr}(M_\ell \not \in g(Y^n)|{\mathcal E}_\ell)\nonumber\\
    &=\mbox {Pr}\left(\bigcup\limits_{\substack{{\sett\in \seto(\ell),}\\{\bar X^n_{\sett}\neq X^n_{\sett}}}}
    \left\{\tilde \gamma(\bar X^n_\sett, X^n_{[\ka]\setminus \sett})>\tilde \gamma(X^n_{[\ka]})\right\}\bigg| {\mathcal E}_\ell\right)\label{apen1: proof thm1; definition of decoder}\\
    &=\mbox {Pr}\left(\bigcup_{\substack{{\sett\in \seto(\ell),}\\{\bar X^n_{\sett}\neq X^n_{\sett}}}}
    \left\{\gamma(\bar X^n_\sett)>\gamma(X^n_{\sett})\right\}\bigg| {\mathcal E}_\ell\right)\label{apen1: proof thm1; chain rule of inf. den.}\\
    &=\mathds{E}\left[\mbox {Pr}\left(\bigcup_{\substack{{\sett\in \seto(\ell),}\\{\bar X^n_{\sett}\neq X^n_{\sett}}}}
    \left\{\gamma(\bar X^n_\sett)>\gamma(X^n_{\sett})\right\}\bigg| X^n_{[\ka]}, Y^n, {\mathcal E}_\ell\right)\right]\label{apen1: proof thm1: random coding}\\
    &\leq\mathds{E}\bigg[\min\bigg\{1, \sum_{\sett\in \seto(\ell)} \binom{\textsfM-\ka}{|\sett|} |\sett|!  \nonumber\\
    &\hspace{2.4cm}  \left. \cdot \mbox {Pr}\left(
    \gamma(\bar X^n_\sett)>\gamma(X^n_{\sett})\bigg| X^n_{[\ka]}, Y^n, {\mathcal E}_\ell\right)\right]\label{apen1: proof thm1; union bound}\\
    &\leq\mathds{E}\bigg[\min\bigg\{1, \sum_{\sett\in \seto(\ell)} \textsfM^{|\sett|} \exp(-\gamma(X^n_{\sett}))\bigg\}\bigg]  \label{apen1: proof thm1; inf. den. bound}\\
    &\leq \sum_{\sett\in \seto(\ell)} \mbox{E}\bigg[\min\bigg\{1, \textsfM^{|\sett|} \exp(-\gamma(X^n_{\sett}))\bigg\}\bigg]  \label{apen1: proof thm1; moving sum out of min}\\
    &\leq \sum_{\sett\in \seto(\ell)} \mbox{Pr}\bigg( \textsfM^{|\sett|} \exp(-\gamma(X^n_{\sett}))\geq U\bigg)  \label{apen1: proof thm1; V>U}\\
    &=\sum_{\sett\in \seto(\ell)} \mbox{Pr}\bigg( \log\Big(\textsfM^{|\sett|} \exp(-\gamma(X^n_{\sett}))\Big)\!-\!\log(U)\!\geq \!0\bigg) \!\!\label{apen1: proof thm1; W tail prob1} \\
    &= \sum_{\sett\in \seto(\ell)} \mbox{Pr}(W_\sett\geq 0),\label{apen1: proof thm1; W tail prob}
    % &=\sum_{\sett\subseteq[\ka], \ell \in \sett} \mbox{Pr}\bigg(\textsfM^{|\sett|} \exp(-\gamma(X^n_{\sett}))>U\bigg) \label{apen1: proof thm1; exp. min to prob.}
\end{flalign}
where (\ref{apen1: proof thm1; definition of decoder}) is due to the definition of the maximum information density decoder, (\ref{apen1: proof thm1; chain rule of inf. den.}) is due to the chain rule of information density. The random coding scheme and union bound are used in (\ref{apen1: proof thm1: random coding}) and (\ref{apen1: proof thm1; union bound}), respectively. Note that the asynchronous model does not have the permutation-invariant property. Therefore, the number of permutations of the erroneously decoded messages, $|\sett|!$, is summed up. The inequality (\ref{apen1: proof thm1; inf. den. bound}) follows from the fact that $\binom{\textsfM-\ka}{|\sett|}\cdot |\sett|!\leq \textsfM^{|\sett|}$ and $$\mbox{Pr}(\gamma(\bar X^n_\sett)>\gamma( X^n_\sett))\leq \exp(-\gamma(X^n_\sett)),$$ where $\bar X^n_\sett$ is an independent copy of $X^n_\sett$\cite[Corollary~18.4]{Polyanskiy_lecture_note}.
The inequality (\ref{apen1: proof thm1; moving sum out of min}) follows from $\min\{1, \beta_1+\beta_2\}\leq \min\{1, \beta_1\}+\min\{1, \beta_2\}$ for $\beta_1, \beta_2\in\mathbb R$. The inequality \eqref{apen1: proof thm1; V>U} follows from $\mathds{E}[\min\{1, V\}]=\mbox{Pr}(V \geq U)$ \cite[eq.(77)]{Polyanskiy.2010} for a non-negative random variable $V$, where $U \sim \mbox{Unif}(0,1)$ is independent of $V$. The equality \eqref{apen1: proof thm1; W tail prob} follows from defining $$W_\sett:=\log\left(\textsfM^{|\sett|} \exp(-\gamma(X_\sett^n))\right)-\log (U).$$
% {\color{red} Sorry, I miss again where is $\ell$ in the 2nd inequality above?}

% For any non-negative random variable $V$ and $U \sim \mbox{Unif}(0,1)$, $\mathds{E}[\min\{1, V\}]$ can be expressed as $\mbox{Pr}(V \geq U)$ \cite[(77)]{Polyanskiy.2010}. Therefore, we can express (\ref{apen1: proof thm1; moving sum out of min}) as $\sum_{\sett\subseteq[\ka], \ell \in \sett} \mbox{Pr}(W_\sett\geq 0)$, where $W_\sett:=\log\left(\textsfM^{|\sett|} \exp(-\gamma(X_\sett^n))\right)-\log (U)$. 
We apply the CGF, the Taylor expansion, and the inverse Laplace transform to derive $\mbox{Pr}(W_\sett\geq 0)$. 
We denote by $\psi_{W_\sett}(t)=\log (\mathds {E}[\exp(t W_\sett)])$ the CGF of the random variable $W_\sett$ with parameter $t$. 
\begin{flalign}
    &\psi_{W_\sett}(t)\nonumber\\
    &= \log \bigg(\mathds {E}\left [\exp\left(t \log\left(\textsfM^{|\sett|} \exp(-\gamma(X_\sett^n))\right)\right.\right.-t \log (U)\Big)\Big]\bigg)\\
     &= t |\sett| \logM-\log(1-t)+\log(\mathds{E}[\exp(-t\cdot \gamma(X^n_{\sett}))])\\
     &= t |\sett| \logM-\log(1-t)-E(\ba,t)\label{apen1: proof thm1: Gaussian integral}\\
     &=\tilde {\psi}_{W_\sett}(t)-\log(1-t),\label{apen1: proof thm1: Gaussian integral2}
\end{flalign}
where $t\in(0,1)$, $\tilde {\psi}_{W_\sett}(t):=t|\sett|\logM-E(\ba,t)$, and (\ref{apen1: proof thm1: Gaussian integral}) is due to the following definition in Theorem 1, 
\begin{align*}
    E(\ba,t)&:=-\log(\mathds{E}[\exp(-t\gamma(X^n_{\sett}))])\\
    &=\frac{1}{2}\sum\limits_{i=1}^{n}\left(t\log(1+a_i \pow)+\log\left(1-\frac{a_i \pow t^2}{1+a_i \pow}\right)\right),
\end{align*} 
where 
\begin{align*}
    \exp(\!-t\!\cdot\!\gamma(X^n_{\sett}))\!\!=\!\prod_{\ell=1}^{n}\!\left(\!\frac{dP_{Y|X_{[\ka]}}\!\left(Y_\ell|\big\{\tau_{d_m}(X^n_{m},\ell)\big\}_{m\in[\ka]}\right)}{dP_{Y}(Y_\ell)}\!\right)^{\!t}\!.
\end{align*}

For $t\in(0,1)$, the CGF converges, which is proved as follows. Since the CGF is the summation of the logarithm of the following $n$ terms, 
\begin{align}
    \mathds{E}[\exp(t \cdot i(\{\tau_{d_m}(X^n_{m},\ell)\}_{m\in\sett}; Y_\ell|\{\tau_{d_m}(X^n_{m},\ell)\}_{m\in\kaml}))], \label{eq:appen: MGF}
\end{align} 
$\ell=1,2,...,n$, for a CGF to converge, a sufficient condition is that (\ref{eq:appen: MGF}) converges in term of $t$ for all $\ell \in [n]$. We apply the Gaussian integral to derive (\ref{eq:appen: MGF}). The corresponding range of convergence for any $\ell \in [n]$ is $t\in\left(-\frac{1+a_\ell\pow}{a_\ell\pow},\sqrt{\frac{1+a_\ell \pow}{a_\ell \pow}}\right)$. When $t<0$, it is possible that $|\sett| \logM > \sum_{\ell=1}^n \frac{1}{2} \log (1+a_\ell \pow)$, which means that the corresponding error probability approaches $1$. For all $\ell \in [n]$ and $a_\ell \in \mathbb Z^+$, $\sqrt{\frac{1+a_\ell \pow}{a_\ell\pow}}\geq 1$. If $a_\ell=0 $, $\ell \in [n]$, it represents that no codeword symbol is transmitted at the $\ell$-th channel use. Thus, the information density is $0$ and the corresponding \eqref{eq:appen: MGF} must converge.
% When $a_\ell=0$, the information density is $0$ and the CGF must converge. Therefore, we apply the function, $\max\{1,a_\ell\}$, to ensure the case $a_\ell$ does not influence the range of $t$. 
Therefore, in Theorem \ref{thm1} and Theorem $\ref{thm2}$, we choose $t=\T\in(0,1)$, which fulfills 
\begin{align}
    |\sett|\logM=E^{(1)}_{t}(\ba,\T), \label{eq:app: sda constriant}
\end{align}
to guarantee the convergence.  

The PDF of $W_\sett$ is obtained by the inverse Laplace transform:
\begin{align}
    f_{W_\sett}(w)=\frac{1}{2\pi j}\int^{c+j\infty}_{c-j\infty}\exp(\psi_{W_\sett}(t)-t w) dt,
\end{align}
where $c\in(0,1).$
The probability, $\mbox{Pr}(W_\sett\geq0)$, is obtained by changing the order of integration, i.e., 
\begin{flalign}
\mbox{Pr}(W_\sett\geq0)&=\frac{1}{2\pi j}\int^{\infty}_0 \left\{\int ^{c+j\infty}_{c-j\infty}\exp(\psi_{W_\sett}(t)-t w) dt\right\}dw\\
    &=\frac{1}{2\pi j}\int ^{c+j\infty}_{c-j\infty}\exp(\psi_{W_\sett}(t)) \frac{dt}{t}\\
    &=\frac{1}{2\pi j}\int ^{c+j\infty}_{c-j\infty}\exp(\tilde {\psi}_{W_\sett}(t)) \frac{dt}{t(1-t)}.\label{apen1: proof thm1: CDF first step}
\end{flalign}
The last equality follows from (\ref{apen1: proof thm1: Gaussian integral2}).
By applying the Taylor expansion to $\tilde{\psi}_{W_\sett}(t)$ at the point $t=\T$, which fulfills \eqref{eq:app: sda constriant}, we have
% recall $\ba$ is a set indicating the number of transmitted symbols that belong to $\sett$ 
\begin{align}
    \tilde{\psi}_{W_\sett}(t)=&\T|\sett|\logM-E(\ba,\T)\nonumber\\
    &+\big[|\sett|\logM-E^{(1)}_{t}(\ba,\T)\big](t-\T)\nonumber\\
    &-E^{(2)}_{t}(\ba,\T)\frac{(t\!-\!\T)^2}{2}\!+\!\bar \xi(\ba,\T,t),\label{apen1: proof thm1: Taylor expansion}
\end{align}
where $\big[|\sett|\logM-E^{(1)}_{t}(\ba,\T)\big](t-\T)=0$ due to \eqref{eq:app: sda constriant},
$$\bar \xi(\ba,\T,t):=\sum_{i=3}^{\infty} -E^{(i)}_t(\ba, \T)\frac{(t-\T)^{i}}{i!}$$ is the sum of higher order terms of Taylor expansion, and $\T$ satisfies (\ref{eq:app: sda constriant}). Substitute (\ref{apen1: proof thm1: Taylor expansion}) and $c=\T$ into (\ref{apen1: proof thm1: CDF first step}), we have
% \begin{strip}
\begin{flalign} 
% (跨行)
    &\frac{1}{2\pi j}\int ^{\T+j\infty}_{\T-j\infty}\exp(\tilde {\psi}_{W_\sett}(t)) \frac{dt}{t(1-t)}\nonumber\\
    &=\frac{\eta}{j}\!\int^{\T\!+j\infty}_{\T-j\infty}\!\!\exp\!\!\bigg(\!\beta\frac{(t\!-\!\T)^2\!}{2}\!+\!\bar\xi(\ba,\T,t)\!\bigg) \frac{dt}{t(1\!-\!t)} \nonumber\\
    &\quad \\
    % &=\!\!\!\int^{\T\!+j\infty}_{\T-j\infty}\!\!\exp\!\!\bigg(\!\!\!-\!\!E^{(2)}_{t}\!(\ba\!,\T)\frac{\!(t\!-\!\T)^2\!}{2} +\bar \xi(\ba,\T)\bigg)\nonumber\\
    % &\quad+\bar \xi(\ba,\T)\bigg) \frac{dt}{t(1-t)} \cdot \frac{g_1(\ba\!,\T)}{2\pi j} \\
    &=\!\!\eta \bigg\{ \frac{1}{j}\int^{\T\!+j\infty}_{\T-j\infty}\!\exp\bigg(\beta\frac{(t\!-\!\T)^2}{2}\bigg) \frac{dt}{t(1\!-\!t)} \nonumber\\
    &\hspace{5cm}  + 2\pi \xi(\ba,\T)\bigg\}\!\!\label{apen1: proof thm1: exponential decomposition}\\
    &\!=\!\eta\bigg\{\!\int ^{\infty}_{-\infty}\!\!\exp\!\bigg(-\beta\frac{\rho^2}{2}\!\bigg) \frac{d\rho}{\T\!\!+\!j\rho} \nonumber\\
    &\quad \!+\! \int ^{\infty}_{-\infty}\!\exp\bigg(\!\!-\beta\frac{\rho^2}{2}\bigg) \frac{d\rho}{1\!-\!\T\!-\!j\rho}\!+\! 2\pi \xi(\ba,\T)\bigg\}, \label{apen1: proof thm1: Before Voigt}
    % &\frac{\eta}{j}\!\!\!\int^{\T\!+j\infty}_{\T-j\infty}\!\!\exp\!\!\bigg(\!\!\!-\!\!E^{(2)}_{t}\!(\ba\!,\T)\frac{\!(t\!-\!\T)^2\!}{2} +\bar \xi(\ba,\T)\bigg) \frac{dt}{t(1-t)} \\
    % &=\!\!\int^{\T+j\infty}_{\T-j\infty}\!\!\exp\!\bigg(\!\!\!-\!E^{(2)}_{t}(\ba\!,\T)\frac{(t\!-\!\T)^2}{2}\bigg) \frac{dt}{t(1\!-\!t)} \nonumber\\
    % &\hspace{0.85cm} \cdot \frac{\eta}{j} + g_1(\ba,\T) \xi(\ba,\T)\label{apen1: proof thm1: exponential decomposition}\\
    % &\!=\!\!\eta\!\bigg\{\!\!\int ^{\infty}_{-\infty}\!\!\exp\!\bigg(\!E^{(2)}_{t}(\ba,\T)\frac{\rho^2}{2}\!\bigg) \frac{d\rho}{\T\!\!+\!j\rho} \nonumber\\
    % &\hspace{0.2cm} + \int ^{\infty}_{-\infty}\exp\bigg(E^{(2)}_{t}(\ba,\T)\frac{\rho^2}{2}\bigg) \frac{d\rho}{(1-(\T+j\rho))}\!\bigg\} \nonumber\\
    % &\hspace{3cm}+ g_1(\ba,\T) \xi(\ba,\T), \label{apen1: proof thm1: Before Voigt}
\end{flalign}
% \end{strip}
where $\eta\!:=\!\frac{g_1(\ba\!,\T)}{2\pi}$, $\beta\!:=\!-E^{(2)}_{t}(\ba,\T)$, $\rho\!:=\!\frac{t-\T}{j}$, and
$$g_1(\ba,t):=\exp(t|\sett|\logM-E(\ba,t)).$$ The equality (\ref{apen1: proof thm1: exponential decomposition}) follows from $e^x=1+\sum_{i=1}^{\infty}\frac{x^i}{i!}$ and by letting $x=\bar \xi(\ba,\T, t)$,
\begin{flalign}
    &\xi(\ba,\T):=\!\frac{1}{2\pi j}\int^{\T+j\infty}_{\T-j \infty}\!\!\exp\left(\frac{\beta}{2}(t\!-\!\T)^2\!\right)\nonumber\\
    &\hspace{3cm} \cdot \frac{1}{t(1-t)} \sum_{m=1}^{\infty}\frac{\bar \xi(\ba,\T,t)^m}{m!} dt.\label{apen1: proof thm1: xi}
\end{flalign}
% We define $\rho:=\frac{t-\T}{j}$ in (\ref{apen1: proof thm1: Before Voigt}).
By multiplying $\frac{\T-j\rho}{\T-j\rho}$ to the first integral in (\ref{apen1: proof thm1: Before Voigt}), we have
\begin{flalign}
    &\int ^{\infty}_{-\infty}\exp\bigg(-\beta\frac{\rho^2}{2}\bigg) \frac{d\rho}{\T+j\rho}\nonumber \\
    &=\int ^{\infty}_{-\infty}\exp\bigg(-\beta\frac{\rho^2}{2}\bigg) \frac{\T d\rho}{\T^2+\rho^2}\nonumber\\
    &\hspace{2cm}- \int ^{\infty}_{-\infty}\exp\bigg(-\beta\frac{\rho^2}{2}\bigg) \frac{j \rho d\rho}{\T^2+\rho^2}\label{apen1: proof thm1: odd func1}\\
    &=\int ^{\infty}_{-\infty}\exp\bigg(-\beta\frac{\rho^2}{2}\bigg) \frac{\T d\rho}{\T^2+\rho^2}\label{apen1: proof thm1: odd func2}\\
    &=2 \pi \exp\left(\frac{\T^2 \beta}{2}\right)Q\left(\T\sqrt{\beta}\right) \label{apen1: proof thm1: apply Voigt func}\\
    &\leq \frac{\sqrt{2 \pi}}{\T}\frac{1}{\sqrt{\beta}}\label{apen1: proof thm1: apply Q func bound2}\\
    &= \frac{\sqrt{2 \pi}}{\T}\frac{1}{\sqrt{-E^{(2)}_{t}(\ba,\T)}},\label{apen1: proof thm1: apply Q func bound1}
\end{flalign}
where the second integral in (\ref{apen1: proof thm1: odd func1}) is the integral of an odd function, which equals $0$. By applying the Voigt function \cite{Finn:1965:TBF} to the integral in (\ref{apen1: proof thm1: odd func2}), we have (\ref{apen1: proof thm1: apply Voigt func}). The inequality (\ref{apen1: proof thm1: apply Q func bound1}) follows from the upper bound of the Gaussian Q-function, $Q(x)\leq \frac{1}{ x \sqrt{2\pi}}\exp(-\frac{x^2}{2})$. The last equality follows from the definition: $\beta\!:=\!-E^{(2)}_{t}(\ba,\T)$.
By multiplying $\frac{1-\T+j\rho}{1-\T+j\rho}$ with the same steps used in deriving (\ref{apen1: proof thm1: apply Q func bound1}), the second integral in (\ref{apen1: proof thm1: Before Voigt}) is bounded by
\begin{flalign}
    &\!\!\!\!\int ^{\infty}_{-\infty}\!\!\exp\bigg(\!\!-\beta\frac{\rho^2}{2}\bigg) \frac{d\rho}{1\!-\!\T\!-\!j\rho}\nonumber\\
&\hspace{2.5cm} \leq \frac{\sqrt{2 \pi}}{1-\T}\frac{1}{\sqrt{-E^{(2)}_{t}(\ba,\T)}}.\!\!\!\label{apen1: proof thm1: apply Q func bound2}
\end{flalign}
Consequently, we can upper bound the sum of the two integrations in (\ref{apen1: proof thm1: Before Voigt}) as follows:
% of (\ref{apen1: proof thm1: Before Voigt}) can be expressed as follows:
\begin{align}
    &\eta\bigg\{\!\int ^{\infty}_{-\infty}\!\!\exp\!\bigg(-\beta\frac{\rho^2}{2}\!\bigg) \frac{d\rho}{\T\!\!+\!j\rho} \nonumber\\
    &\quad \!+\! \int ^{\infty}_{-\infty}\!\exp\bigg(\!\!-\beta\frac{\rho^2}{2}\bigg) \frac{d\rho}{1\!-\!\T\!-\!j\rho}\!+\! 2\pi \xi(\ba,\T)\bigg\}\nonumber\\
    &\leq \frac{g_1(\ba,\T)}{\sqrt{2\pi}(1-\T)\T}\frac{1}{\sqrt{-E^{(2)}_{t}(\ba,\T)}}\nonumber\\
    &\hspace{3.1cm}+g_1(\ba,\T) \xi(\ba,\T).\label{apen1: proof thm1: final}
\end{align}
By combining (\ref{apen1: proof thm1; split to no-repetition event}), (\ref{apen1: proof thm1; W tail prob}), (\ref{apen1: proof thm1: CDF first step}), (\ref{apen1: proof thm1: Before Voigt}), and (\ref{apen1: proof thm1: final}), the PUPE of the AUMAC system for a given $\setd$ is 
\begin{align}
    &\sum_{\ell=1}^{\ka}\frac{1}{\ka}\mbox {Pr}(M_\ell \not \in g(Y^n)|\setd, {\mathcal E}_\ell)+p_0\nonumber\\
    % &\leq \sum_{\ell=1}^{\ka}\frac{1}{\ka}\sum_{\sett\subseteq[\ka], \ell \in \sett} \mbox{E}\bigg[\min\bigg\{1, \textsfM^{|\sett|} \exp(-\gamma(X^n_{\sett}))\bigg\}\bigg]+p_0\\
    &\leq \sum_{\ell=1}^{\ka}\!\frac{1}{\ka}\!\sum_{\substack{\sett\in\seto(\ell)}}\!\!\bigg\{\!\frac{g_1(\ba,\T)}{(1\!-\!\T)\T}\frac{1}{\sqrt{\!-2\pi E^{(2)}_{t}\!(\ba,\T)}}\nonumber\\
    &\hspace{2.3cm} + g_1(\ba,\T) \xi(\ba,\T)\bigg\}+p_0\\
    &= \sum_{\sett\subseteq [\ka]}\frac{|\sett|}{\ka}\bigg\{\frac{g_1(\ba,\T)}{(1-\T)\T}\frac{1}{\sqrt{-2 \pi E^{(2)}_{t}(\ba,\T)}}\nonumber\\
    &\hspace{2.2cm} + g_1(\ba,\T) \xi(\ba,\T)\bigg\}+p_0,\label{apen1: proof thm1: final2}
\end{align}
where $\seto(\ell)$ is defined in \eqref{app: eq: def of seto}.

%%%%%%%%%%%

\section{Proof of Theorem 2}\label{app:sec:thm2 proof}
In the following, in addition to Theorem \ref{thm1}, we derive a uniform upper bound of the PUPE of an $(n, \textsfM, \epsilon, \ka, \alpha)$-code
% w.r.t. the blocklength $n$, message size $\textsfM$, PUPE constraint $\epsilon$, the number of active users $\ka$, and the delay constraint $\alpha n$, 
as indicated in Theorem 2. In particular, we will find the worst-case asynchronicity, which implies finding the worst-case of $\ba$ and $\T$ in Theorem \ref{thm1}.
% The formula of PUPE in Theorem 1 can be treated as a function with two inputs, $\ba$ and $t$. 
To simplify the derivation, we denote $\iota:= \ind(1\in\sett)$ and all possible $\ba$'s w.r.t. $\iota$ by the set
$\mathcal F_{k,\iota}:=\{ \ba: \iota, |\sett|=k\}$, where $\ba$ is defined in \eqref{eq:def a} as a function of $\sett$ and $\setd$.
% We also denote $\mathcal F:= \bigcup_{k\in[\ka]} \{\mathcal F_{|\sett|,0}\cup\mathcal F_{|\sett|,1}\}$.

We will show that for all $t\in (0,1)$, there exists an $a_\iota^{n*}$ resulting in a uniform upper bound of PUPE for all $\ba \in \mathcal F_{|\sett|,\iota} $, such that the upper bound of the PUPE in \eqref{eqthm1: main result} has the following property 
\begin{flalign*}
    &g_1(\ba\!,t)g_2(\ba\!,t)\leq \!g_1(a_\iota^{n*}\!,t)g_2(a_\iota^{n*}\!,t).
\end{flalign*}

However, for any $\ba \!\in\! \mathcal F_{|\sett|,\iota}$, the order between $$g_1(\ba\!,\T)g_2(\ba\!,\T)$$ and $$g_1(a_\iota^{n*}\!,t_0(a_\iota^{n*}))g_2(a_\iota^{n*}\!,t_0(a_\iota^{n*}))$$ is not fixed, 
since the sign of $\frac{\partial}{\partial t} g_1(\ba,t)g_2(\ba,t)$ is not the same for all $t\in(0,1)$.
Therefore, for fixed $a_\iota^{n*}$, we will show that the choices of $T^*_0$, $T^*_1$, $\underline t_0$, and $\underline t_1$ uniformly upper bound the PUPE regardless $\setd$. 
% Since we do not bound the term $p_0$ and also the approximation error term $\xi(\ba,\T)$, we omit them in the following derivation.
We start from \eqref{eqthm1: main result} restated as follows
\begin{align}
\sum_{\sett\subseteq[\ka]}\frac{|\sett|}{\ka\sqrt{2\pi}}  g_1(\ba,\T) g_2(\ba,\T),
\end{align}
while omitting the term $p_0$ and also the approximation error term $\xi(\ba,\T)$ since we do not bound these terms.

To proceed, we use the following lemma.
\begin{lemma}
\label{lemma: the product of 2 functions are non-decreasing}
    Let $g_1(\ba,t)=\exp(f_1(\ba,t))$ and $g_2(\ba,t)=(f_2(\ba,t))^{-\frac{1}{2}}$, where $\ba\in \{\mathbb Z^+_ 0\}^n$, $t\in(0,1)$, $f_1(\ba,t)\in\mathbb R$ and $f_2(\ba,t)> 0$. Then $g_1(\ba,t)g_2(\ba,t)$ is a non-increasing function w.r.t. $a_i,\,\forall i\!\in\![n]$ if $f_{1,a_i}^{(1)}(a_i,t)\!\leq\! 0$ and $f_{2,a_i}^{(1)}(a_i,t)\!\geq\! 0$.
\end{lemma}
The proof of Lemma 1 is relegated to Appendix \ref{app:proof: lemma}.

Then, we apply Lemma \ref{lemma: the product of 2 functions are non-decreasing} by defining $f_1(\ba,t):=t|\sett|\logM-E(\ba,t)$ and $f_2(\ba,t):=-(t-t^2)^2E_t^{(2)}(\ba,t)$. 
% We first calculate the derivative: $ \frac{d g_1(\ba,t) g_2(\ba,t)}{da_i}= g^{(1)}_{1,a_i}(a_i,t)g_2(\ba,t)+g_1(\ba,t) g^{(1)}_{2,a_i}(a_i,t)$ regarding $a_i,\;i\in[\alpha a]$. If $g^{(1)}_{1,a_i}(a_i,t)\leq0$ and $g^{(1)}_{2,a_i}(a_i,t)\leq0$, $\frac{d g_1(\ba,t) g_2(\ba,t)}{da_i}\leq 0$ since $g_1(\ba,t)$ and $g_2(\ba,t)$ are positive. 
% Equivalently, we can check the signs of the first derivative of the exponent of $g_1(\ba,t)$ and the denominator of $g_2(\ba,t)$, respectively. We define $f_1(\ba,t):=\log(g_1(\ba,t))=t|\sett|\logM-E(\ba,t)$ and $f_2(\ba,t):=(g_2(\ba,t))^{-2}=-(t-t^2)^2E^{(2)}_t(\ba,t)$.
The first derivatives of $f_{1}(\ba,t)$ and $f_2(\ba,t)$ w.r.t. $a_i$ are expressed as follows, respectively 
\begin{align}\label{app: f_1 partial derivative a}
    f_{1,a_i}^{(1)}(a_i,t)=&\frac{\pow (t^2-t)+a_i\pow^2(t^3-t)}{2(1+a_i\pow)(1+a_i\pow-a_i\pow t^2)}
\end{align}
and
    \begin{align}\label{app: f_2 partial derivative a}
    f_{2,a_i}^{(1)}(a_i,t)=&\frac{\pow (1-t)^2t^2(1+a_i \pow +3 a_i \pow t^2)}{(1+a_i \pow -a_i \pow t^2)^3}.
\end{align}

For $t\in(0,1)$, it is clear that $f^{(1)}_{1,a_i}(a_i,t)\leq0$ and $f^{(1)}_{2,a_i}(a_i,t)\geq 0$. We then conclude that $g_1(\ba,t)g_2(\ba,t)$ is a non-increasing function w.r.t. $a_i,\;i\in[\alpha n]$ according to Lemma \ref{lemma: the product of 2 functions are non-decreasing}. It implies that the PUPE of any given $\sett$ decreases with increasing $a_i,\;i\in[\alpha n]$. Namely, reducing $a_i,\;i\in[\alpha n]$ will upper bound the error probability. 
% We conclude that $g_1(\ba,\T) g_2(\ba,\T)$ is a non-increasing function with respect to $a_i,\;i\in[\alpha n]$, i.e., the PUPE of any given $\sett$ decreases as $a_i,\;i\in[\alpha n]$, increases. Namely, reducing the value of $a_i,\;i\in[\alpha n]$ upper bounds the error probability. 
Therefore, to upper bound the PUPE, we can consider the following case, where the number of transmitted symbols that belong to $\sett$ at the first $\alpha n$ channel use, $a_{[\alpha n]}$, are reduced to the minimum, which is $a^*_{\iota,[\alpha n]}=\iota$. Namely $a^{n*}_\iota=[\iota^{\alpha n}, |\sett|^{n-\alpha n}]$. Consequently, for all $\ba\in{\mathcal F}_{|\sett|,\iota}$ and a given $t=\T$, we have 
\begin{align}
    &g_1(\ba,\T)g_2(\ba,\T)\nonumber\\
    &\hspace{2.5cm}\leq g_1(a^{n*}_\iota,\T)g_2(a^{n*}_\iota,\T).\label{eq: Appen B: g1 1}
\end{align}

We have shown that the error probability is non-increasing w.r.t. $a_i$. 
However, the sign of $\frac{\partial }{\partial t} g_1(\ba, t)g_2(\ba, t)$ w.r.t. $t$ changes for $t\in(0,1)$.
To solve it, we can show that given $a^{n*}_\iota$, if $t_0(a^{n*}_\iota)\in\mathcal A\cap \mathcal B$, $\bar {t}_\iota \in \mathcal A \cap \bar {\mathcal B}$, $\underline {t}_\iota\in \mathcal A \cap \underline {\mathcal B}$, and $\underline {t}_\iota\leq \T\leq\bar {t}_\iota$, then there exist a uniform upper bound of the error probability for all $\setd$ satisfying delay constraint $\alpha n$, where
% $\ba\in\mathcal F_{|\sett|,\iota}$ and a given $a^{n*}_\iota$, where  
\begin{flalign}
        &\mathcal {A}\!:=\left\{t\!:\!t\in(0,1) \right\},\\
        &\mathcal B\!:=\left\{t\!:\!E^{(1)}_t(a^{n*}_\iota, t)=|\sett|\logM\right\},\label{eq:app:set: B}\\
        % &=\left\{t:\sum_{i=1}^n \frac{a^*_{\iota,i}\pow t}{1+a^*_{\iota,i}\pow-a^*_{\iota,i}\pow t^2}=\frac{1}{2}\sum_{i=1}^{n}\log(1+a^*_{\iota,i}\pow)-|\sett|\logM\right\}\\
        &\bar {\mathcal B}\!:=\left\{t\!:\!\sum_{i=1}^{n}\frac{a^*_{\iota,i}\pow t}{1\!+\!a^*_{\iota,i}\pow(1\!-\! t^2)}\!=\!\frac{n}{2} \log(1\!+\!|\sett|\pow)\!-\!|\sett|\logM\right\},\label{def:bar t}\\
        &\underline {\mathcal B}\!:=\left\{t\!:\!\frac{|\sett| n\pow t}{1\!+\!|\sett|\pow\!-\!|\sett|\pow t^2}\!=\!\sum_{i=1}^{n}\frac{1}{2}\log(1\!\!+\!a^*_{\iota,i}\pow)\!-\!|\sett|\logM\right\},\label{def:underline t}
        % &\text{and } a^{*}_{\iota,i} \text{ is the } i \text{-th element of } a^{n*}_\iota.\nonumber
\end{flalign}
and $a^*_{\iota,i}$ is the $i$-th element of $a^{n*}_\iota$.

To proceed, we find upper bounds of $g_1(a^{n*}_\iota,\T)$ and $g_2(a^{n*}_\iota,\T)$ as $u_1$ and $u_2$. Then we upper bound $g_1(a^{n*}_\iota,\T)g_2(a^{n*}_\iota,\T)$ by $u_1u_2$.
% We will show that $T_0(a^{n*}_\iota)$ results in a uniform upper bound of $g_1(a^{n*}_\iota,t)$ and the choice of $\underline t_\iota$ and $\bar t_\iota$ result in the uniform upper bound of $g_2(\ba,t)$ for all $\ba \in \mathcal F_{|\sett|,\iota}$.
Since the second partial derivative w.r.t. $t$,
\begin{align}
    f^{(2)}_{1,t}(\ba,t)=\sum_{i=1}^{n}\frac{(a_i\pow)+(a_i\pow)^2+(a_i\pow)^2t^2}{(1+a_i\pow-a_i\pow t^2)^2},
\end{align}
is positive for $t\in(0,1)$, $f_1(\ba,t)$ is a convex function regarding $t$. Moreover, $f^{(1)}_{1,t}(\ba,\T)=0$ by \eqref{eq:app:set: B}. Namely, $f_1(\ba,t)$ achieves minimum at $t=\T$. Therefore, for any $\ba \in \mathcal F_{|\sett|,\iota}$, we have
\begin{align}
    g_1(\ba,\T)\leq g_1(\ba,t_0(a^{n*}_\iota))\leq g_1(a^{n*}_\iota,t_0(a^{n*}_\iota)),\label{eq:app:thm2 proof: g1}
\end{align}
where the first inequality is because $g_1(\ba,t)$ achieves the minimum at $t=\T$. If $\ba=a^{n*}_\iota$, the equalities hold. The second inequality follows from the fact that $g_{1}(\ba,t)$ is a non-decreasing function for a given $t\in(0,1)$ w.r.t. $a_i,\;\forall i\in[\alpha n]$, since $$g_{1,a_i}^{(1)}(a_i,t)=\exp(f_1(\ba,t))\cdot f_{1,a_i}^{(1)}(a_i,t)\leq 0,$$ for $t\in(0,1)$, where $f_{1,a_i}^{(1)}(a_i,t)$ is given in (\ref{app: f_1 partial derivative a}).
% \color{red} and $a^{n*}_\iota=[\iota^{\alpha n}, |\sett|^{n-\alpha n}]$ reduces $a^*_{i}$ to the minimum, $i\in[\alpha n]$.\color{black}
% follows (\ref{eq: Appen B: g1 1}) and (\ref{eq: Appen B: g1 2}) for $\iota=1$ and $\iota=0$, respectively.

We define $f_2(\ba,t):=(f_3(t))^2 f_4(\ba,t)$, where $f_3(t):=t-t^2$ and $f_4(\ba,t):=-E^{(2)}_t(\ba,t)$.
% There exist $\underline t_\iota$ and $\bar t_\iota$ for any $\ba \in \mathcal F_{|\sett|,\iota}$, $\iota=\{0,1\}$ such that $$\underline t_\iota\leq \T\leq \bar t_\iota$$.
Since the first partial derivative w.r.t. $t$ of $f_{4,t}(\ba,t)$ is as follows
\begin{align}
   f^{(1)}_{4,t}(\ba,t)= \sum_{i=1}^{n} 2(a_i\pow)^2t \frac{3+3 a_i \pow +a_i \pow t^2}{(1+a_i \pow -a_i \pow t^2)^3},
\end{align}
which is positive for $t\in(0,1)$, $f_4(\ba,t)$ is a non-decreasing function of $t$. 
% \color{red}Additionally, for all $\ba\in\mathcal F_{|\sett|,\iota}$, there exists $\underline t_\iota \leq \T$. 
% Because $f_4(\ba,t)$ is a non-decreasing function w.r.t. $t$ and , 
Then, we have 
\begin{align}
    f_4(a^{n*}_\iota,t_0(a^{n*}_\iota))\geq f_4(a^{n*}_\iota, \underline t_{\iota}).\label{eq:app:proof:g2:f4}
\end{align}
% Therefore, we can conclude that 
% \begin{align}
% \frac{1}{f_4(a^{n*}_\iota,\T)}\leq \frac{1}{f_4(a^{n*}_\iota, \underline t)}. \label{eq: app: f4 final}
% \end{align}
% (It seems $\bar t$ just appear but no reason)
\color{black} 
% In addition, since $f_3(t):=t-t^2$ is a concave function  
% \begin{align}
%    f_3(\T)\geq& \lambda f_3(\underline t_\iota)+\bar \lambda f_3(\bar t_\iota)\\
%     \geq &\lambda \min\{f_3(\bar t_\iota), f_3(\underline t_\iota)\}+\bar \lambda\min\{f_3(\bar t_\iota), f_3(\underline t_\iota)\}\\
%     =&T^*_\iota,
% \end{align}
% where $\lambda\in [0,1]$. Let $t_1=\bar t_\iota$, $t_2=\underline t_\iota$. There exists a $\lambda$ such that $\lambda t_1+(1-\lambda) t_2=\T$. 

By the condition, $\underline t_\iota\leq \T\leq \bar t_\iota$, there exists a $\lambda\in[0,1]$ such that $\T=\lambda\underline t_\iota+\bar \lambda\bar t_\iota$, where $\bar \lambda=1-\lambda$. Since $f_3(t)$ is concave, it satisfies 
\begin{flalign}
f_3(\T)\geq& \lambda f_3(\underline t_\iota)+\bar \lambda f_3(\bar t_\iota)\\
    \geq &\lambda \min\{f_3(\bar t_\iota), f_3(\underline t_\iota)\}+\bar \lambda\min\{f_3(\bar t_\iota), f_3(\underline t_\iota)\}\\
    =&\min\{f_3(\bar t_\iota), f_3(\underline t_\iota)\} =: T^*_\iota. \label{eq:app:Proof thm2 :T*}
\end{flalign}
% where $T^*_\iota:=\min\{f_3(\bar t_\iota), f_3(\underline t_\iota)\}$.

Consequently, for all $\ba\in \mathcal F_{|\sett|,\iota}$, we have   
\begin{align}
g_1(a^{n*}_\iota,\T)\leq g_1(a^{n*}_\iota,t_0(a^{n*}_\iota)),
\end{align}
stated in \eqref{eq:app:thm2 proof: g1}, and 
\begin{align}
g_2(a^{n*}_\iota,\T)&=\frac{1}{f_3(\T) \sqrt{f_4(a^{n*}_\iota,\T)}} \label{eq:app:proof thm2 :final1}\\
&\leq \frac{1}{T^*_\iota \sqrt{f_4(a^{n*}_\iota,\T)}}\label{eq:app:proof thm2 :final2}\\
&\leq \frac{1}{T^*_\iota \sqrt{f_4(a^{n*}_\iota,\underline t_\iota)}},\label{eq:app:proof thm2 :final3}
\end{align}
where \eqref{eq:app:proof thm2 :final1} is by definition, \eqref{eq:app:proof thm2 :final2} follows from \eqref{eq:app:Proof thm2 :T*} and \eqref{eq:app:proof thm2 :final3} follows from \eqref{eq:app:proof:g2:f4}. 

Consequently, we have 
\begin{align}
   &g_1(\ba,\T)g_2(\ba,\T)\nonumber\\
   &\leq g_1(a^{n*}_\iota,\T)g_2(a^{n*}_\iota,\T)\\
   &\leq g_1(a^{n*}_\iota,t_0(a^{n*}_\iota))g_2(a^{n*}_\iota,\T)\\
   % &=g_1(a^{n*}_\iota,T_0(a^{n*}_\iota))\frac{1}{f_3(\T) \sqrt{f_4(a^{n*}_\iota,\T)}}\\
   &\leq \frac{g_1(a^{n*}_\iota,t_0(a^{n*}_\iota))}{T^*_\iota\sqrt{-E^{(2)}_t(a^{n*}_\iota, \underline t_\iota)}},
\end{align}
which completes the proof of Theorem 2.
% Combining (\ref{proofthm2: f1}), (\ref{proofthm2: g2}), and indicator functions with new condition, $\{\bar t_\iota,\;\underline t_\iota \in(0,1),\;\forall \ba^*_\iota\in \mathcal F\}$.

% $the semi-convexity and $

\section{Proof of Lemma \ref{lemma: the product of 2 functions are non-decreasing}}\label{app:proof: lemma}
Let $g_1(\ba,t)=\exp(f_1(\ba,t))$ and $g_2(\ba,t)=(f_2(\ba,t))^{-\frac{1}{2}}$, where $\ba\in \{\mathbb Z^+_0\}^n$, $t\in(0,1)$ and $f_1(\ba,t)\in \mathbb R$, and $f_2(\ba,t)>0$. Then the first partial derivative of $g_1(\ba,t)g_2(\ba,t)$ w.r.t. $a_i$ is 
% a non-increasing function w.r.t. $a$ if $f_{1,a_i}^{(1)}(a_i,t)\leq 0$ and $f_{2,a_i}^{(1)}(a_i,t)\geq 0$.
\begin{align}
    &\frac{\partial}{\partial a_i} g_1(\ba,t)g_2(\ba,t)\nonumber\\
    &=g_2(\ba,t) \frac{\partial}{\partial a_i} g_1(\ba,t) +g_1(\ba,t) \frac{\partial}{\partial a_i} g_2(\ba,t)\\
    % &=g_2(\ba,t) \frac{\partial}{\partial a_i} \exp(f_1(\ba,t))+g_1(\ba,t)\frac{\partial}{\partial a_i}(f_2(ba,t))^{-\frac{1}{2}}\\
    &=g_2(\ba,t) \exp(f_1(\ba,t)) \frac{\partial}{\partial a_i} f_1(\ba,t)\nonumber\\
    &\hspace{1.5cm}-\frac{1}{2} g_1(\ba,t) (f_2(\ba,t))^{-\frac{3}{2}}\frac{\partial}{\partial a_i}f_2(\ba,t).
\end{align}
Therefore, $g_1(\ba,t)g_2(\ba,t)$ is a non-increasing function w.r.t. $a_i$, if $\frac{\partial}{\partial a_i} f_1(\ba,t)\leq 0$ and $\frac{\partial}{\partial a_i}f_2(\ba,t) \geq 0$.

%%

%%
%% If several appendices are needed, then the command
%%
% \appendices
%%
%% in combination with further \section-commands can be used.
%%%%%%

\end{document}